\documentclass[superscriptaddress,aps,pra,twocolumn,preprintnumbers,amsmath,amssymb,floatfix,groupedaddress]{revtex4}

\usepackage[dvips]{graphicx}
\usepackage{dcolumn}
\usepackage{bm}
\usepackage{amssymb}
\usepackage{xr}

\usepackage{color}

\newcommand{\cst}[1]{\ensuremath{\mathrm{#1}}} 
\newcommand{\e}{\cst e}

\newcommand{\eop}{\mathcal{E}}

\begin{document}

\title{Controlling the quantum state of a single photon emitted from a single
polariton }

\author{ Jovica Stanojevic, Valentina Parigi, Erwan Bimbard,  Rosa
Tualle-Brouri, Alexei Ourjoumtsev and Philippe Grangier}
\affiliation{Laboratoire Charles Fabry, Institut d'Optique, CNRS, Universit\'e
Paris-Sud, Campus Polytechnique, RD 128, 91127 Palaiseau cedex, France }

\date{\today}


\begin{abstract}
We investigate in detail the optimal conditions for a high fidelity transfer from a
single polariton state to a single photon state and subsequent homodyne detection of the single photon.
We assume that, using various possible techniques, the single polariton has initially been stored as a spin-wave grating in a cloud of cold atoms inside a low-finesse cavity. This state is then transferred to a single photon optical pulse using an auxiliary beam. We optimize the retrieval efficiency and determine the mode of the local oscillator that maximizes the homodyne efficiency of such a photon. We find that both efficiencies can have values close to one in a large region of experimental parameters.
\end{abstract}


\maketitle

\section{Introduction}

The generation and characterization of  well-controlled single photon states have
accomplished considerable progress during recent years.
Beyond the usual non-classical effects obtained using single  photon counting
and intensity correlation measurements \cite{Glauber}, interference effects
offer very efficient ways to get more information about the quality of
single-photon wave packets. For instance, observing coalescence, that is the
famous ``Hong-Ou-Mandel" effect \cite{HOM}, between two single photons emitted
by  two different atomic sources is a very good way to ensure that these two
photons are indistinguishable \cite{Beugnon}.

Another avenue in order to fully characterize a single photon state is using
homodyne detection, and applying  quantum tomography techniques in order  to
reconstruct the Wigner function $W(x,p)$ of the photon in phase space, where
$\hat x$ and $\hat p$ are the quadrature operators of the quantized electric
field \cite{Hansen}. Such a method provides a full characterization of the
quantum state of the single photon, which is very intuitive because the quality
of the single photon state, and especially its purity, directly translate into
the negativity of the Wigner function.

In the optical domain quantum homodyne tomography has been implemented for
various number states, including one-  \cite{Hansen} and two-  \cite{Ourj2,Val}
photon Fock states. In these experiments, the usual method to produce a single-photon state is to use
parametric fluorescence emitting non-degenerate pairs of photons, and to detect
one of them using a photon counter, such as an avalanche photodiode. Then the
other beam is projected in the desired one-photon Fock state, which is analyzed
by homodyne detection and then quantum tomography. It has
to be noted that the quantum efficiency of the homodyne detection channel has to
be very high: any loss at that stage degrades the purity of the single photon
state by mixing it with vacuum, and quickly destroy the negativity of the Wigner
function.  This method works quite well, but it is intrinsically
non-deterministic~: the probability of the first (photon counting) event must be
low, and then the photon cannot be emitted ``on demand", when needed for
applications.

\begin{figure}
\includegraphics[scale = 0.3]{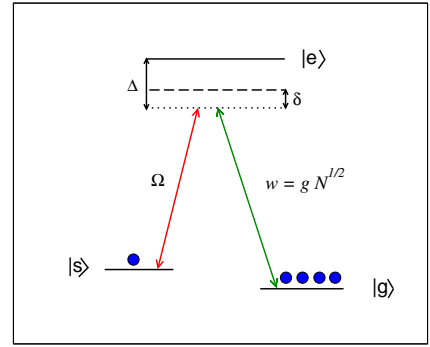}
\caption{(Color  online)
An ensemble of $N$ three-level atoms placed  in a cavity is initially prepared
in a single polariton state, with $1$ atom in the metastable state $| s \rangle$
and $(N\!-\!1)$
atoms in the ground state $| g \rangle$.
The atoms are coupled to the quantum field with  collectively enhanced coupling
$w\!=\!g \sqrt{N}$, where $g$ is the single photon Rabi frequency. By applying a
classical control field (red) with Rabi
frequency $\Omega(t)$ a single-photon state is generated with well defined
properties.  The cavity detuning $\delta$ is used to optimize the photon
retrieval efficiency.
\label{fig:scheme}}
\end{figure}

Here we would like to investigate another scheme, which can, at least in
principle, be made more deterministic. The idea is to emit the single photon
from a so-called ``polariton" state, where a single long-lived spin excitation is
distributed over many atoms. 
It can be converted into a 
free-propagating single photon using a control laser beam (see Fig.~\ref{fig:scheme}). 
As we will show below, given the presence of the polariton the source is deterministic: 
the phase-matching condition between the readout laser beam, the atomic spin wave 
and the single photon leads to a collective enhancement effect which ensures that 
the photon is emitted in a well-defined spatial and temporal mode. For the whole scheme 
to be deterministic, the polariton should also be prepared in a deterministic way, 
which can be done for instance by using the
so-called ``Rydberg blockade" mechanism as explained in  recent  theoretical
papers \cite{blockade,1phot}. However, even when the preparation of the polariton is probabilistic, 
one may exploit the fact that the single photon can be emitted on demand, just when it is actually needed, within
the relatively long coherence time of the spin wave.
In this article we will not consider in more details how the polariton is
prepared \cite{Gorshkov07,Dantan04,Dantan06,Gorshkov08,Kalachev08}, 
but rather look at the following question: assuming that a single
polariton is imprinted on an atomic cloud within an optical cavity, how
efficiently can we turn it back into a single photon within a single mode? What
will be the amplitude and frequency of the emitted single photon wave-packet,
and is it possible to control it, and to mode-match it efficiently onto a local
oscillator? In order to answer these questions, we will use as a starting point
the theoretical analysis presented in
\cite{Gorshkov07}, which has been shown to be in good agreement with experiments
\cite{Vuletic}.  We will extend  this analysis in order to characterize very
carefully the homodyne efficiency which can be obtained for such a single-photon
pulse, since this efficiency is crucial for quantum homodyne tomography. Finally
we will discuss various experimental and theoretical considerations.

\section{Theory}
We assume that the atoms are confined in an optical cavity and  a single
polariton state has been prepared so we only analyze the retrieval of a single
photon. We are primarily interested in situations in which the collectively
enhanced coupling between the atoms and a quantized cavity radiation mode is
always one of the dominant energy scales, but not necessarily the only one.
This is an interesting regime for experiments and it has not been studied in
detail before. The underlying atom-cavity model has been already described
\cite{Gorshkov07} and thus we only briefly introduce it.

\subsection{Cavity model}

Various techniques and strategies for photon storage and  retrieval have been
analyzed using a relatively simple model \cite{Gorshkov07}. It is assumed
that a cloud of $N$ three-level  atoms
(see Fig.~\ref{fig:scheme}) with two metastable states $\left|g\right>$ and
$\left|s\right>$, and an excited state  $\left|e\right>$ are coupled to a
classical and  quantized single-mode fields. The classical (read) field couples
the states $\left|s\right>$ and $\left|e\right>$ while the quantized field
couples the states
$\left|g\right>$ and $\left|e\right>$. The coupling to the classical field is
characterized by a slowly varying Rabi frequency $\Omega(t)$ and the
collectively enhanced coupling $w$  between the atoms and the quantized cavity
radiation mode is $w\!=\!g\sqrt{N}$, where $g$ is the single photon Rabi frequency.

The assumption that, initially, a single excitation is shared by a very large
number of atoms practically means that almost all atoms are in the ground state
at all times so that the following approximate relations are justified:
$\Sigma_{i=1}^N\sigma^i_{gg}\!\approx\!N$, and
$\Sigma_{i=1}^N\sigma^i_{ee}\!\approx\!\Sigma_{i=1}^N\sigma^i_{ss}
\!\approx\!\Sigma_{i=1}^N\sigma^i_{es}\!\approx\!0$, where
$\sigma^i_{uv}\!=\!\left|u\right>_{i\,i}\!\left<v\right|$.
Under these simplifications and in the rotating-wave approximation,  we define
the atomic polarization annihilation operator
$\hat P(t)\!=\!\sum_{i=1}^N\sigma^i_{ge}/\sqrt{N}$, and spin-wave annihilation
operator $\hat S(t)\!=\!\sum_{i=1}^N\sigma^i_{gs}/\sqrt{N}$.
In the general case, these operators also include  phase factors
$\exp(-i \;  {\bf k.r_i})$, where $\bf k$ is the laser wave-vector, and $\bf
r_i$ the position of atom number $i$.
There phase factors can be absorbed by redefining the state $\left|g\right>_i$.
In Appendix A we write the Langevin equations for
the coupled propagation of  the field annihilation operator $\hat \eop(t)$ and
atomic operators $\hat P(t)$ and $\hat S(t)$  in the Heisenberg picture,
and we show that it is actually enough to consider the c-number equations~:
\begin{eqnarray}
\dot {\eop} &=& -(\kappa + i \delta) \;  \eop + i w P
\label{cav_eqs1},\\
\dot {P} &=& - (\gamma+i \Delta) \; P + i w  \eop + i \Omega S ,
\label{cav_eqs2}\\
\dot {S} &=&  i \Omega^*  P ,\label{cav_eqs3}
\end{eqnarray}
where $\kappa$ and $\gamma$ are respectively the cavity and polarization decay
rates, while $\delta$ is the detuning between the  laser-driven Raman light
and the cavity, whereas $\Delta$  the detuning between the driving laser and the atomic line (see Fig.  \ref{fig:scheme}).

It should be pointed out   that the c-numbers $\eop(t)$, $P(t)$ and $S(t)$
are {\bf not}  averages of the corresponding operators, but can be seen as
quantum amplitudes, which allow one to  evaluate
normally ordered products of operators (see details in Appendix A).
An intuitive justification for this approach is that, since
there is only one excitation in the system, there will be only one emitted
photon in the cavity mode; we are therefore calculating the amplitude of this
single photon wave packet.  The initial conditions   are
$\eop(t_0)\!=\!P(t_0)\!=\!0$ and $S(t_0)\!=\!1$, where $t_0$ indicates the
beginning of the retrieval sequence.

The retrieval efficiency $\eta$ is defined as the ratio of retrieved photons and
stored excitations.
Since the number of stored excitations is assumed to be one, $\eta$ is equal to
the number of retrieved photons
\begin{equation} \label{eta}
\eta \!=\! \int_{t_0}^\infty d t |\eop_\textrm{\rm out}(t)|^2,
\end{equation}
where
\begin{equation} \label{io}
\eop_\textrm{\rm out}(t) = \sqrt{2 \kappa} \; \eop(t).
\end{equation}
As said before, we do not need to use quantum operators in Eq. (\ref{eta}) since
we never have more than one photon in the quantized radiation mode.

The propagation equations for $(\eop_R,P_I,S_R)$ and  $(\eop_I,P_R,S_I)$
decouple for  $\Delta\!=\!\delta\!=\!0$ and real $\Omega$, where $a_R$ and $a_I$ are the
real and imaginary part of a complex  variable $a$. Consequently,  $\eop$ and
$S$ are real and $P$ is imaginary in that case.
For real $\Omega$, there is the following mapping between the solutions for~$\pm
\Delta$
\begin{eqnarray}
\eop(-\Delta,-\delta,t) &=&\eop^*(\Delta,\delta,t) \label{map1},\\
P(-\Delta,-\delta,t) &=& -P^*(\Delta,\delta,t),\label{map2}\\
S(-\Delta,-\delta,t) &=&  S^*(\Delta,\delta,t) .\label{map3}
\end{eqnarray}
For non-zero $\Delta$, the interaction between the atoms and the quantized
radiation mode shifts the cavity resonance. Therefore, to compensate this shift
and maximize the efficiency (\ref{eta}),
we introduce some cavity detuning $\delta\!\neq\!0$.  Relations
(\ref{map1})-(\ref{map3}) impose  the mapping between the optimized cavity
detunings $\delta_{\rm opt}$ as follows
\begin{equation} \label{map_delta}
\delta_{\rm opt}(-\Delta)\!=\!-\delta_{\rm opt}(\Delta).
\end{equation}
  As explained in Section \ref{section:HD-optimization}, other mappings relevant
for the optimization of homodyne detection (HD) also follow from Eqs.
(\ref{map1})-(\ref{map3}). All these mappings  also have practical usage since
they reduce numerical work by half.

\subsection{Retrieval efficiency}\label{section:theory}

We want to explore the parameter space by means of  numerical calculations to
find regions of high photon retrieval efficiency.
In many experiments the emission of the single photon is detected by a counter,
such as an avalanche photodiode, and the relevant parameter is $\eta$ as defined
by Eqs. \ref{eta} and \ref{io}.
Here, we are also interested in using homodyne detection, which allows us to
analyze fully the single photon wavepacket, and e.g. to reconstruct its Wigner
function by using quantum optical tomography.  Therefore we will also look for
optimal conditions for homodyne detection.

Let us first review the expression for the maximal photon retrieval efficiency
in a general case since our search for optimal retrieval conditions do not
necessarily involve any  special cases such as the bad-cavity limit ($\kappa\gg
w$). Actually, based on experimental considerations, the typical range of
parameters we are interested in
corresponds to $w >> \kappa \sim \gamma$. We will show now that the efficiency
maximum is then the  same one derived in
 \cite{Gorshkov07}, though we are no more in the bad cavity limit.

A common way to simplify the problem is to try to find the first integrals of
the equations of motion. Since our system is dissipative
($\kappa,\gamma\neq0$), we may at least look for some relations analogous to the
first integrals. The following relation gives just that
\begin{equation} \label{first_int}
\frac{d}{d t}\left( |\eop|^2+ |P|^2 + |S|^2\right) = - 2 \gamma  |P|^2   - 2
\kappa  |\eop|^2.
\end{equation}

We are interested in the conditions for high retrieval efficiencies $\eta$.
Using Eqs. (\ref{io}) and (\ref{first_int}), as well as the conditions
$S(t_0)\!=\!1$,
$\eop(t_0)\!=\!\eop(\infty)\!=\!P(t_0)\!=\!P(\infty)\!=\!S(\infty)=0$, one gets
the following relation
\begin{equation}\label{int_form}
1 =\eta+ 2 \gamma \int_{t_0}^{\infty} |P(t)|^2 dt .
\end{equation}
The boundary condition $S(\infty)\!=\!0$ is  only fulfilled for maximal
efficiencies since the spin-wave decay rate is neglected.
It is convenient to  rewrite the integral in the last equation in the frequency
space and then use the relation between the Fourier components $P(\omega)$ and
 $\eop(\omega)$ obtained from Eq. (\ref{cav_eqs1})
\begin{equation} \label{ft_cav_eqs1}
 P(\omega) =\frac{\delta-\omega-i \kappa}{w} \eop(\omega) .
\end{equation}
Therefore
\begin{eqnarray}
\int\limits_{t_0}^{\infty} |P(t)|^2 dt &\! =\!& \int\limits_{\infty}^{\infty}
|P(\omega)|^2 d\omega  \nonumber \\
&\! =\!&\frac{\kappa^2}{w^2}\!\int\limits_{-\infty}^{\infty}
\left[1\!+\!\frac{(\omega-\delta)^2}{\kappa^2}\right]|\eop(\omega)|^2
d\omega,\label{P_int}
\end{eqnarray}
which implies
\begin{equation} \label{ft_cav_eqs2}
2\gamma\int\limits_{t_0}^{\infty} |P(t)|^2 dt \!\geqslant\!   \frac{\kappa
\gamma}{w^2}
\left[ 2\kappa\int\limits_{t_0}^{\infty} |\eop(t)|^2 dt \right]\!=\! \frac1C
\eta,
\end{equation}
where $C\!=\!w^2/\kappa \gamma$ is the cooperativity factor. Combining this
condition and Eq. (\ref{int_form}), we get the efficiency limit
\begin{equation} \label{eff_limit}
\etaÊ\; \!\leqslant\! \frac{C}{1+C}.
\end{equation}
As shown in  \cite{Gorshkov07} for $\delta\!=\!0$, it is possible  to achieve
the highest  efficiency $C/(1\!+\!C)$ in some particular regimes. Since the relation (\ref{eff_limit}) 
is very general, introducing an optimized cavity detuning $\delta_{\rm opt}\!\neq\! 0$ may allow us to optimize 
the efficiency, especially for low intensities of the driving field, but it will never exceed this limit.

A necessary, but not sufficient, condition to have a maximal efficiency is
that  the term $(\omega-\delta)^2/\kappa^2$ in Eq. (\ref{P_int}) should be
negligible. For $\delta\!=\!0$, this happens if the range of frequencies
$\omega$ corresponding to non-vanishing $P(\omega)$ and $\eop(\omega)$ is much
smaller than $\kappa$. In this case, achieving high efficiencies would require
long read pulses or high $\Omega$. However, according to Eq.
(\ref{P_int}), the efficiency can be  improved significantly by introducing a cavity detuning
$\delta$ that cancels  $\omega$ corresponding to the maximum of
$|\eop(\omega)|$. A consequence is that $\delta_{\rm opt}$ is comparable with
the spectral width of
$|P(\omega)|$ and $|\eop(\omega)|$ so that we expect that both
$\omega^2/\kappa^2$ and $\delta_{\rm opt}^2/\kappa^2$ are negligible for very
long pulses.
  Because the term $\omega^2/\kappa^2$ in Eq. (\ref{P_int}) originates from
$\dot{\eop}(t)$ in  Eq. (\ref{cav_eqs1}), we may think that  $\dot{\eop}(t)$
can be neglected  for very high efficiencies. Dropping this
term yields to the `bad cavity' relation between $\eop(t)$ and $\hat P(t)$
\begin{equation} \label{bad_cav}
-\kappa  \eop + i w  P =0,
\end{equation}
and therefore ($\kappa,w>0$)
\begin{equation} \label{bad_cav_abs-val}
\kappa | \eop| =  w | P|.
\end{equation}
In general, as shown in detail in Appendix B, for sufficiently long pulses and high efficiencies a 
better approximation can be obtained by noting that Eq. (\ref{cav_eqs1}) is the first-order expansion of
\begin{equation} \label{bad_cav-corr}
-\kappa \eop(t+\delta t) + i w P(t) =0,
\end{equation}
with respect to the short constant time delay $\delta t=1/\kappa$. Equation (\ref{bad_cav-corr}) 
substitutes to Eq. (\ref{bad_cav}) in the time domain, while in the frequency domain it becomes
\begin{equation} \label{bad_cav-corrFreq}
-e^{i\phi_{\omega}}\kappa\eop(\omega)+i w  P(\omega) =0
\end{equation}
with $\phi_{\omega}=-\omega \delta t$, and the relationship (\ref{bad_cav_abs-val}) remains valid.

\subsection{Solution for large $w$}\label{section:solution}

In this section, we assume that $w$ is at least large but $\Omega$ and  $\Delta$
can be large as well. For $w \gg \kappa, \; \delta$, relation (\ref{ft_cav_eqs1})
shows that  the ratio $|P|/|\eop|$  is small as long as  the pulsewidths of $\Omega$,
$\eop$, and $P$ are of the same order of magnitude
\begin{equation} \label{PbyE}
\frac{|P|}{|\eop|} \ll 1,
\end{equation}

 Because $S\sim 1$ at short times (due to the initial condition $S(t_0)=1$), Eqs. (\ref{PbyE})
and (\ref{cav_eqs2}) lead to the following estimate of $\eop$ for small
$\Delta$,
\begin{equation} \label{E.vs.W}
\eop\sim -\frac{ \Omega(t)}{w}.
\end{equation}

Let us introduce a new variable $Q=\eop+p P$ and take $p$ for which the following
equation is held
\begin{equation} \label{Q_def}
\dot{Q}=\lambda Q+i p \Omega S.
\end{equation}
There are two combinations of $p,\lambda$ for which the last equation is
fulfilled
\begin{eqnarray}
\lambda_{\pm} =\!&\! -\frac12\left[\gamma\!+\!i\Delta\!+\!k\pm i\sqrt{4 w^2\!
-\!(\gamma\!+\!i\Delta-k)^2} \right],\label{alpha_def} \\
p_{\pm} =\!&\! \frac{i}{2w}\left[\gamma\!+\!i\Delta\!-\!k\pm i\sqrt{4 w^2
\!-\!(\gamma\!+\!i\Delta\!-\!k)^2} \right], \label{a_def}
\end{eqnarray}
where $k=\kappa+i\delta$.
We choose $p$, $\lambda$
which facilitate the adiabatic elimination of $Q$ in Eq. (\ref{Q_def})
\begin{equation} \label{Q_adiab}
\eop+p P=-\frac{i p}{\lambda} \Omega S\equiv-\frac{w}{\beta^2} \Omega S,
\end{equation}
where
\begin{equation} \label{beta_def1}
\beta^2 =-iw\frac{\lambda}{p}
\end{equation}
 is introduced for convenience.

\begin{figure}
\includegraphics[width=9cm]{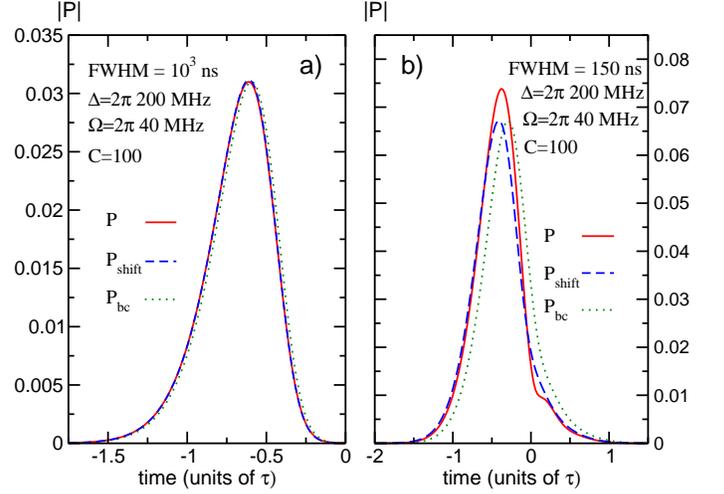}
\caption{(Color  online)
Time delay between $P(t)$ and  $\eop(t)$. (a) For high retrieval efficiencies and
small $\dot{\Omega}/\Omega$  (i.e long pulses), the cavity field  $\eop$ is
delayed with respect to $P$  by a constant delay  $\delta t=1/\kappa$:
 $\eop(t)\approx i w P(t-\delta t) /\kappa$.
The  numerical solution for $P$ (solid red) is compared with  $P_{\rm
bd}(t)\equiv-i \kappa \eop(t)/ w  $ (solid green)  and   $P_{\rm
shift}(t)=P_{\rm bd}(t+\delta t)$ (dashed blue). (b) For shorter pulses ($\tau=150$ ns)
this simple delay approximation breaks and $P(t)$ and  $\eop(t)$ acquire different shapes.
\label{fig:time_delay}}
\end{figure}

Estimates (\ref{PbyE}) and (\ref{E.vs.W}) show that $|P|$ is a much  smaller
quantity compared to both $|\eop|$ and $|S|$. Therefore, dropping the small term
$p P$ in Eq. (\ref{Q_adiab}) has  little effect to the relationship between much
larger $\eop$ and $S$
\begin{equation} \label{E.vs.S}
\eop=-\frac{w}{\beta^2} \Omega S  .
\end{equation}
The last equation can be obtained in a different way and somewhat different
approximations. For some parameters, we can use both $\lambda_{\pm}$ and
$p_{\pm}$ to eliminate $P$ in Eq. (\ref{Q_adiab} ) and get relation
(\ref{E.vs.S}) again  with a new definition for $\beta$
\begin{equation} \label{beta_def2}
\beta^2 =-i w\frac{\lambda_{+}\lambda_{-}}{\lambda_{+}-\lambda_{-}}
\frac{p_{+}-p_{-}}{p_{+}p_{-}}.
\end{equation}
For our typical range of parameters, there is no practical advantage of using
either Eq. (\ref{beta_def1}) or Eq. (\ref{beta_def2}) but it turns out that the
definition (\ref{beta_def2}) is more suitable for very long pulsewidths, as shown 
on Fig. \ref{fig:time_delay}.
Since $S\approx1$  in the beginning of a pulse sequence, we see that, for
initial times, the shape of $|\eop|$ and $|\Omega|$ are always the same and
$|\eop|\ll|\Omega|$.
Taking the derivative of Eq. (\ref{E.vs.S}) with respect to $t$ and using Eqs.
(\ref{cav_eqs1}) and (\ref{cav_eqs3}) to eliminate $\dot S$ and $P$, we get
\begin{equation} \label{E-DE}
\left[1+\frac{|\Omega|^2}{\beta^2}\right]\dot{\eop}=\left[\frac{\dot\Omega}{
\Omega}-k\frac{|\Omega|^2}{\beta^2}\right]\eop.
\end{equation}
For real  $\Omega$, the solution of this equation consistent with the initial
condition $\eop(t_0)=0$ is relatively simple
\begin{equation} \label{E-solut}
\eop(t)= -\frac{w}{\beta} \; \frac{\Omega}{\sqrt{\beta^2+\Omega^2}} \exp\left[-k
\int_{t_0}^{t} \frac{\Omega^2}{\beta^2+\Omega^2} dt'   \right]  .
\end{equation}
where the multiplicative constant $(-w/\beta)$ has been determined by taking
limits of Eqs. (\ref{E-solut}) and (\ref{E.vs.S}) at  $t\rightarrow t_0$ and
utilizing the initial condition $S(t_0)=1$.
Similarly, the solution for $S$  is
\begin{equation} \label{S-solut}
S(t)=\frac{\beta}{\sqrt{\beta^2+\Omega^2}} \exp\left[-k \int_{t_0}^{t}
\frac{\Omega^2}{\beta^2+\Omega^2} dt'   \right]  .
\end{equation}
Finally, the solution for P in our approximation is
\begin{equation} \label{P-solut}
P(t)=\frac{i \beta(k\Omega+\dot \Omega)}{\left(\beta^2+\Omega^2\right)^{3/2}}
\exp\left[-k \int_{t_0}^{t} \frac{\Omega^2}{\beta^2+\Omega^2} dt'   \right]  .
\end{equation}

Let us notice that if $|\Omega|^2\ll|\beta|^2$, then $\eop(t)\approx -w
\Omega(t)/\beta^2=-ip\Omega(t)/\lambda$, which means that $\Omega$ and $\eop$
have the same pulse shape. However, according to Eq. (\ref{E.vs.S}) and
(\ref{S-solut}), this also means that $S\approx1$ and hence the retrieval
efficiency is very low.

\begin{figure}
\begin{center}
\includegraphics[scale = 0.3]{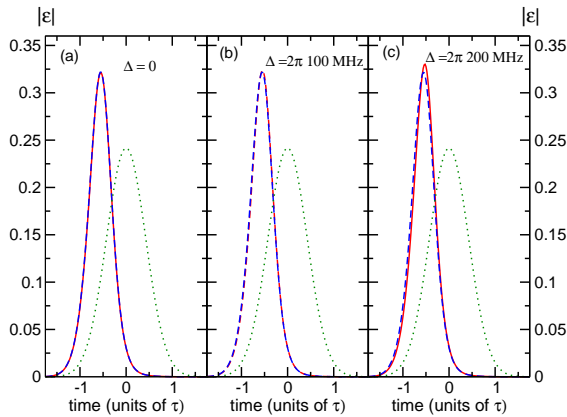}
\end{center}
\caption{(Color  online)
Comparison between numerical (solid red) and analytical (dashed blue) solutions  for
the emitted signal field $|\eop(t)|$.
The Rabi frequency $\Omega(t)$ of the driving field  is  also shown (dashed green), scaled down to match the pulse area of the
corresponding cavity field $|\eop(t)|$. Let us note that
 the (very small) emitted pulse lies entirely within the envelope of the (very large) driving pulse,
 but its peak value does occur before the driving field peak value.
The parameters are $\Omega/2\pi \!=\!80$ MHz, $C\!=\!200$,  $\delta\!=\!0$ MHz, and pulsewidth (FWHM)
$\tau\!=\!200$ ns for several detunings. The agreement is practically exact for
$\Delta\!=\!0$, but it gradually worsens as $\Delta$ increases.
\label{fig:lambda_comparison}}
\end{figure}

Solutions (\ref{E-solut})-(\ref{P-solut}) are excellent approximations for large
$w$ and $\Delta\approx0$, they work rather well for moderate to large $\Delta$,
but for large detuning they become inaccurate, as illustrated in Fig.
\ref{fig:lambda_comparison}. The reason why the approximation worsens depends on
which definition of $\beta$ is used.
For $\Delta=\delta = 0$, both $|\lambda_{\pm}|$  are the same but they can be
very different for larger $\Delta$.  The same statement holds for $|p_{\pm}|$.
Since the larger $|p|$ corresponds to the larger $|\lambda|$, dropping the term
$p P$ in Eq. (\ref{Q_adiab}) becomes problematic for larger $\Delta$ even though
the adiabatic elimination itself works better.  On the other hand, the problem
with using Eq. (\ref{beta_def2})  for larger $\Delta$ is that the adiabatic
elimination corresponding to the smaller $|\lambda|$  becomes unreliable.

\section{Optimization of homodyne detection }\label{section:HD-optimization}

Homodyning consists in interfering two light fields: the local oscillator (LO),
and the measured ``quantum'' beam (or signal beam),
associated with mode annihilation operator $\hat a(t)$.
The  local oscillator is assumed to be  monochromatic
and to be in a coherent state $|\alpha\rangle$, with $\alpha=|\alpha|\e^{-i\theta}$
where $|\alpha|^2$ is the LO photon flux and $\theta$ is a controllable LO phase.
More precisely, the signal from the homodyne detection (HD)  is obtained by
taking the difference of the
two output photocurrents, resulting from the interference of the signal and LO.
This difference photocurrent $i(t)$ is proportional to the following  operator
\begin{equation} \label{HD_signal}
i(t) \propto  \left( \alpha^*\; {\hat a}(t)+\alpha \;  {\hat a}^{\dagger}(t) \right)  / |\alpha|  =  \e^{i\theta} {\hat a}(t)
 + \e^{-i\theta} {\hat a}^{\dagger}(t)
\end{equation}
which is a quadrature operator with respect to the signal field, determined by
the choice of $\theta$.

In the present work, we will consider a situation with a continuous single-mode
LO, but where the signal beam corresponds to a single photon, which is an
essentially transient field, triggered by the ``read-out" pulse sent onto the
atoms in the cavity. This situation is somehow similar to the homodyne detection
of transient homodyne signals following a photodetection event, as measured for
instance in \cite{polzik,sasaki}. The idea is then to ``extract'' from the
stationnary noise coming out from the homodyne detection a non-stationnary
signal, through an appropriate time-domain filtering,  which has to be matched
as well as possible to the expected ``single-photon wavepacket''.

Two effects can essentially degrade this matching~: the first one is the
efficiency $\eta$ introduced in the previous section, i.e. the probability that
the photon is effectively emitted in the output mode $\hat a(t)$, and the second
one is the quality of the interference between the photon and the LO,
represented by an overlap factor $\chi$ defined below.

In order to examine the effect of the quality of the interference, we will
neglect all technical mode-matching issues, and consider only the effect of the
interference of the LO with the single-photon amplitude ${\cal E}(t)$ introduced
in the previous section. Our goal is to optimally match the mode ${\cal E}_{det}(t)$
measured by the homodyne detector to the mode of the single photon defined by
${\cal E}(t)$. The amplitude $f(t)=|{\cal E}_{det}(t)|$ can be controlled either by
modulating the intensity of the local oscillator or by rapidly sampling the signal
and applying a numerical filter $f(t)$ to the data, and in the following we will
assume that the amplitudes of the two modes are identical:
\begin{equation} \label{f-def}
f(t)= \frac{\left| \eop_{\rm out}(t) \right|} {\sqrt{\eta}}.
\end{equation}

The phase $\theta(t)$ of ${\cal E}_{det}(t)$ includes a constant phase shift $\varphi_0$,
which can be introduced on the local oscillator using a piezoelectric actuator,
and correspond to various choices for  the measured quadrature.
In the present situation of  a single photon state,  we don't expect this phase to have any effect
on the  measured signal. In general a
rotating phase $\omega_0 t$ can be added by changing the LO frequency by $\omega_0$.
More complex functions $\theta(t)$ can be realized in principle, for example by chirping
the local oscillator frequency, but for simplicity we will assume that $\theta(t)=\varphi_0+\omega_0 t$.

The mode-matching efficiency can be optimized by maximizing
the variance of the homodyne detection output, integrated over the output pulse.
Up to a multiplicative constant this variance is given by the expression
\begin{eqnarray} \label{HD_variance}
V &=& \left <  \left[ \int dt f(t) i(t) \right] ^2   \right> \\ \nonumber
&=&   \int dt dt' f(t) f(t') \left < i(t) i(t') \right>  \\ \nonumber
&=&   \int dt dt' f(t) f(t') \times \\ \nonumber
&&( \left < \hat a(t) \hat a^\dagger (t') \right> \e^{i \theta(t)-i\theta(t')} +
 \left < \hat a^\dagger (t) \hat a(t') \right> \e^{i \theta(t')-i\theta(t)} + \\ \nonumber
&&  \left < \hat a(t) \hat a(t') \right> \e^{i \theta(t)+i\theta(t')} +
 \left < \hat a^\dagger (t) \hat a^\dagger (t') \right> \e^{-i \theta(t)-i\theta(t')} )
\end{eqnarray}

In the present case, the initial conditions are such that the output cavity field will be a diagonal mixture
of zero and one photon,
and therefore the average values $\left< \hat a(t) \hat a(t') \right> $ and
$ \left < \hat a^\dagger (t) \hat a^\dagger (t') \right>$
are zero for all times (see discussion in Appendix A).
Since the calculated quantities provide normally ordered averages, one uses the commutation relations to get finally~:
\begin{eqnarray}
V &=& 1 + 2   \int dt dt' f(t) f(t')  \left < \hat a^\dagger (t) \hat a(t') \right> \e^{i \theta(t')-i\theta(t)} \\ \nonumber
&=& 1 + 2 \;  I_0
\end{eqnarray}
where
\begin{equation} \label{omega_opt}
I_0= \left| \int dt f(t) \eop_{\rm out}(t)  e^{-i \omega_0 t}\right|^2.
\end{equation}
As expected $V=1$ (vacuum noise) if $I_0=0$, and $V=3$ (single photon state) if $I_0=1$.
It should be noticed that  the phase $\varphi_0$ does not appear any more
(as it could be expected), and
one has thus only to  optimize $V$ with respect to $\omega_0$. The optimal
$\omega_0$, denoted by  $\omega_{\rm opt}$, is obtained by maximizing  the
integral $I_0$ and is  found numerically.
Up to a constant prefactor, the optimized detector signal $I_{\rm max}$ is equal
to the maximal value of $I_0$.
Interpreting the integral in Eq. (\ref{omega_opt}) as a scalar product of $f(t)
e^{i\omega_0 t}$ and $\eop_{out}(t) $, and using the Cauchy-Schwarz inequality for
real and normalized
($\rVert f\rVert\!=\!1$) filters, we get
\begin{equation} \label{I0-ineq}
I_0 \; \leq \;   \rVert\eop \rVert^2=  \int dt  |\eop_{\rm out}(t)|^2 = \eta.
\end{equation}
The upper bound $I_{\rm upper}\!=\! \eta$ of $I_0$, given by Eq.
(\ref{I0-ineq}), can be used to define
a measure $\chi$ of the optimization success, by the expression
\begin{equation} \label{chi-def}
\chi\!= \!\frac{{\rm max}(I_{0})}{I_{\rm upper}}\!=\!\frac{\left|\int dt
f(t) \eop_{\rm out}(t)  e^{-i \omega_{\rm opt} t } \right|^2}
{\eta}.
\end{equation}
The mapping (\ref{map1}) implies the following mapping of the optimized
frequency shift
\begin{eqnarray}
\omega_{\rm opt}(-\Delta) &=& -\omega_{\rm opt}(\Delta) , \label{map_omg}
\end{eqnarray}
\begin{figure}
\includegraphics[scale = 0.35]{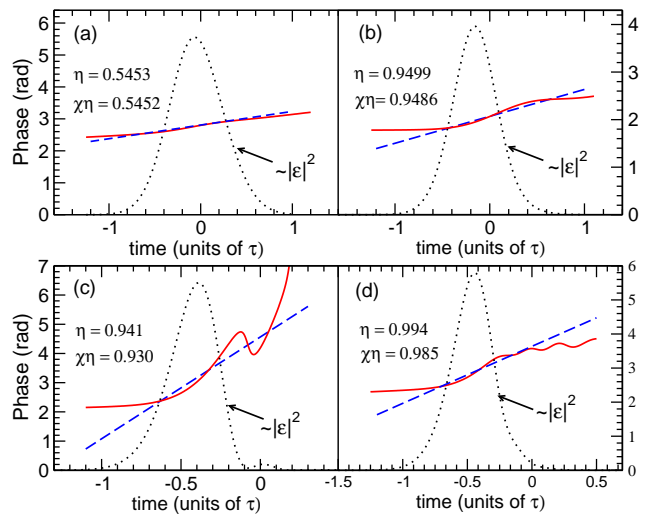}
\caption{(Color  online) 
Linearization of the  phase $\theta_{\eop}$ of the quantum field $\eop(t)$.
To get the highest homodyne efficiency, the phases of the quantum field and local oscillator 
have to be matched (see Section III). Assuming the phase  $\theta_{\rm LO}$  of the local oscillator 
to be linear in time, the optimal frequency $\omega_{\rm opt}$ 
 of the LO and the optimal cavity detuning $\delta_{\rm opt}$  that 
maximize the homodyne efficiency $\chi \eta $ are numerically found
for the generic  parameters 
 $\tau\!=\!150$ ns and $C\!=\!100$ used in section \ref{section:numerix}.  
In panels a) and b), $\theta_{\eop}$ (red solid) and $\theta_{\rm LO}$  (blue dashed) 
are shown for
$\Delta/2\pi \!=\!120$ MHz and $\Omega/2\pi\!=\! 13$ MHz. 
No cavity detuning optimization ($\delta\!=\!0$) is performed in case a) while
an optimized $\delta_{\rm opt}/2\pi\, \!=\!15.5$ MHz is used in case b).
The homodyne efficiency $\chi$ starts to decrease  for  a given $\Delta$ and
sufficiently large $\Omega$ because $\theta_{\eop}$ becomes more complicated, as
shown in panel c)   for $\Omega/2\pi\!=\!80$ MHz 
and $\Delta/2\pi \!=\!
300$ MHz, with $\delta_{\rm opt}/2\pi\, \!=\!-2.8$ MHz. Panel d) is the same as c), except for the cooperativity that becomes C=200,
improving the phase matching quality.
 \label{fig:phase} }
\end{figure}
Nearly maximally allowed value $\eta$ of $I_0$ is achieved by using
the filtering function $f(t)$, and by looking for the best possible linear (in time)
approximation of the phase  of  $\eop_{\rm out}(t)$ (see Fig. \ref{fig:phase}).
As shown below, the regions of high values of $\eta$ and $\chi$ have a very good
overlap, which is important information for future experiments.
Let us emphasize however that in practice, $\eop_{\rm out}(t)$ in Eq.
(\ref{chi-def}) has to be corrected by  including experimental losses, which will result in lower
overall detection efficiencies.
\begin{figure}[t]
\includegraphics[scale = 0.28]{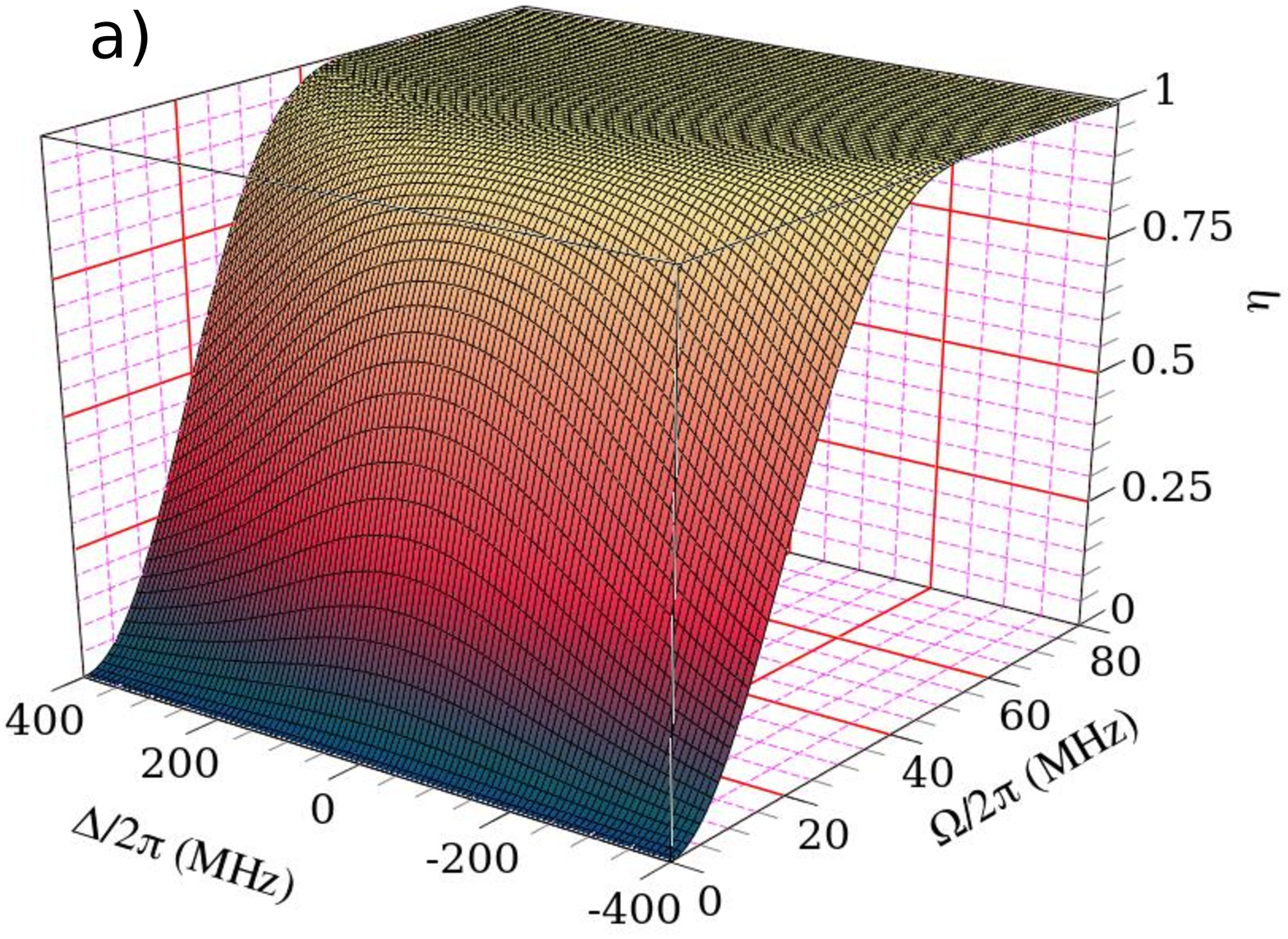}
\includegraphics[scale = 0.28]{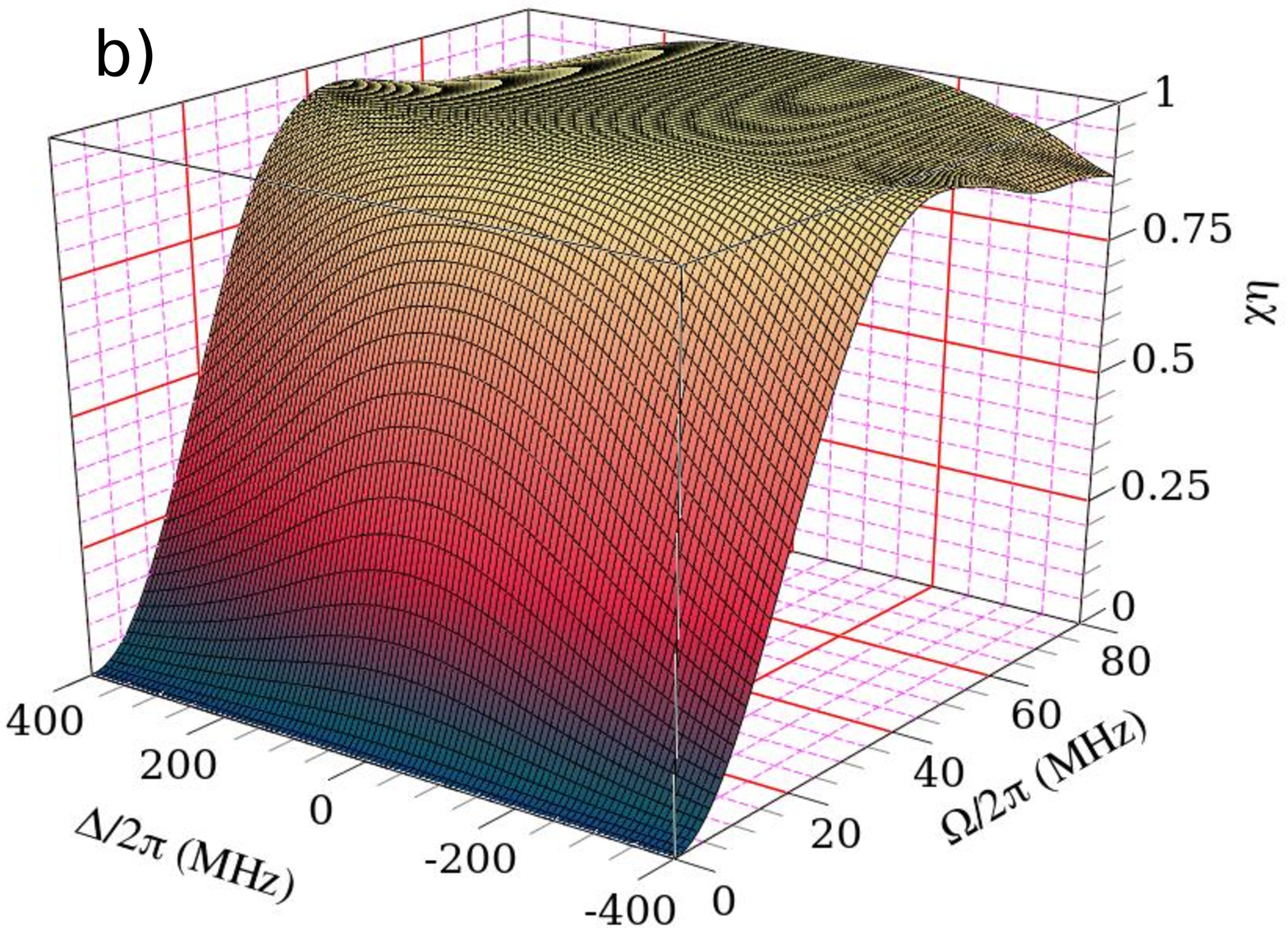}
\caption{(Color  online)
Retrieval efficiency $\eta$ (panel a) and overall homodyne efficiency  $\chi \eta$ (panel b) for non-optimized
cavity detuning (i.e.  $\delta\!=\!0$) for $C\!=\!100$ and $\tau=150$ ns.
The frequency shift $\omega_0$ is 
 optimized to get the best  $\chi \eta$ for any given  $\Delta$ and $\Omega$.
\label{fig:zero_delta}
}
\end{figure}

\section{Results}\label{section:numerix}

In this section we present numerical results on optimal conditions for efficient
retrieval and subsequent detection of a single photon. The efficiencies 
$\eta$ and $\chi$ are evaluated using the numerical solution for $\eop(t)$, 
assuming a readout pulse with a Gaussian shape (as a function of time). 
Based on a typical experimental design using Rb atoms in a cavity \cite{oldQND},
we will take $\kappa/2\pi\!=\!9$ MHz and $\gamma/2\pi\!=\!3$ MHz as fixed
values, and explore a parameter space consisting of $(\Delta,\Omega)$ points.
Similarly, we will assume that the cooperativity factor is in the range
$C\sim100$ \cite{oldQND}, using for instance the two hyperfine states
$\left| g\right>\!=\!\left|5s_{1/2},F\!=\!2,M_F\!=\!2\right>$ and  $\left|
s\right>\!=\!\left|5s_{1/2},F\!=\!1,M_F\!=\!1\right>$ 
of ${}^{87}$Rb. The readout pulse duration is chosen 
short enough (150 ns) so that  the motional decoherence can be neglected
for a cold atom cloud \cite{Vuletic,Vuletic2}. 
\begin{figure}[t]
\includegraphics[scale = 0.28]{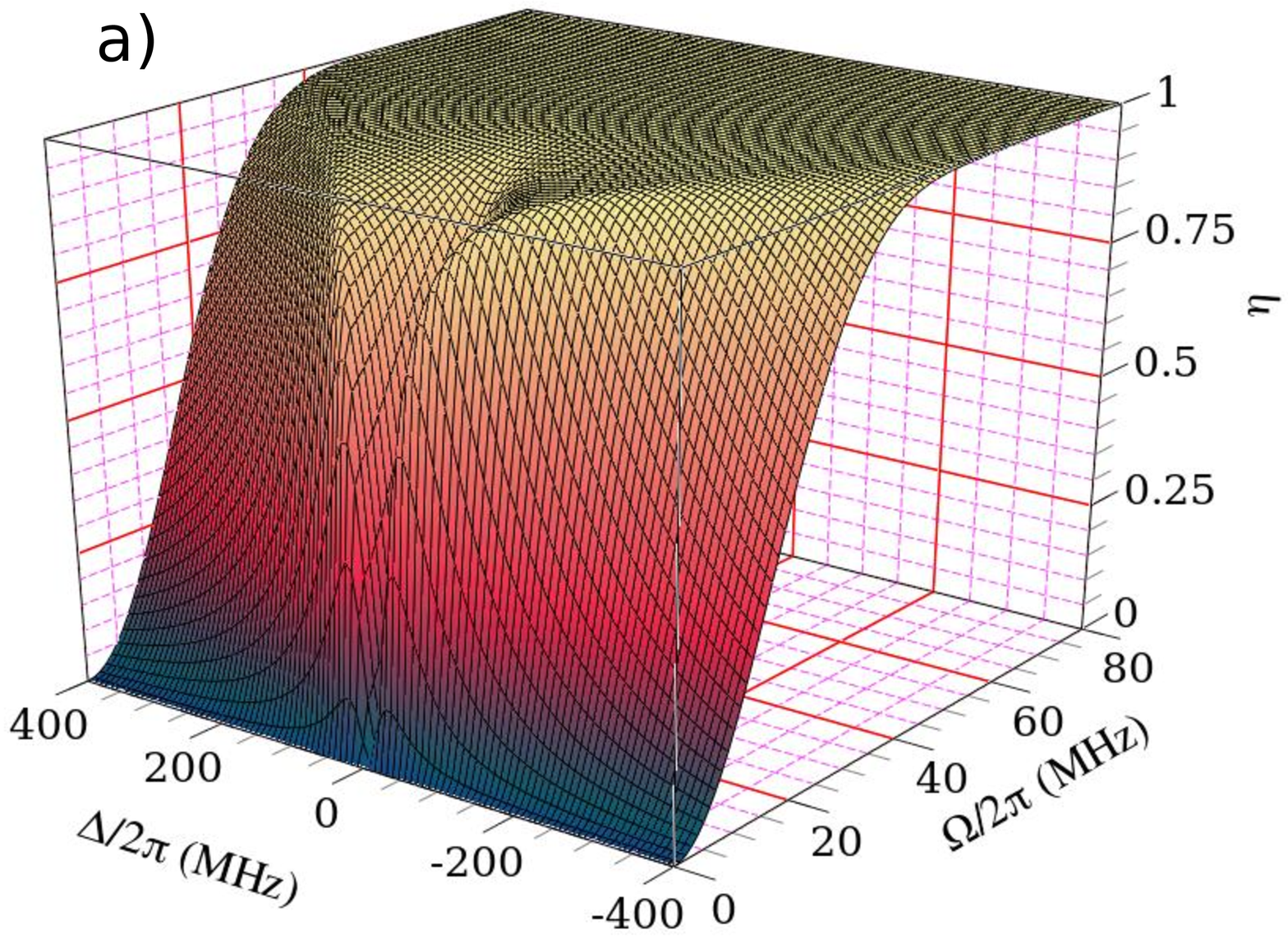}
\includegraphics[scale = 0.28]{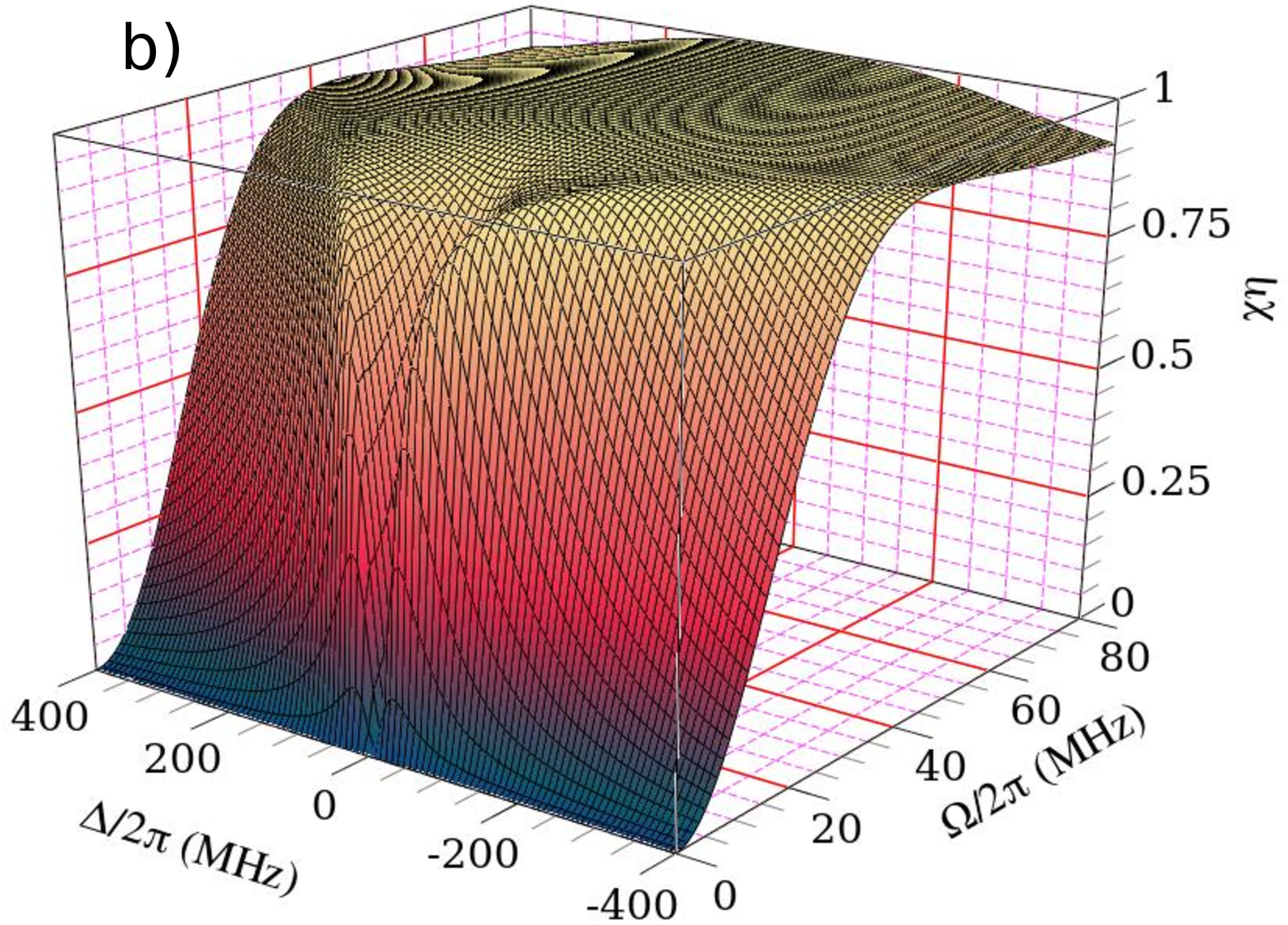}
\caption{(Color  online)
Retrieval efficiency $\eta$ (panel a) and overall homodyne efficiency  $\chi \eta$ (panel b) for optimized
cavity detuning. The efficiency $\chi \eta$  is optimized with respect to  frequency and cavity detuning
(same  parameters  as in Fig. \ref{fig:zero_delta}).
 \label{fig:optim_eta}
}
\end{figure}

\subsection{Optimization of the cavity detuning}

\begin{figure}
\includegraphics[scale = 0.30]{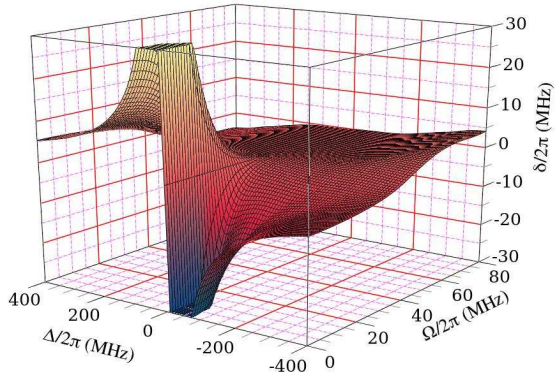}
\caption{(Color  online)
Optimized  cavity detuning  for $\tau\!=\!150$ ns and cooperativity
$C\!=\!100$, which corresponds to $w/2\pi$= 46.5 MHz.
The cavity detuning $\delta_{\rm opt}$ that maximizes the efficiency
$\chi \eta$ is found
for any given $(\Delta,\Omega)$.
 \label{fig:optim_delta}
}
\end{figure}

As pointed out in Section \ref{section:theory}, if  $\Delta \neq 0$ the
frequency of the cavity mode can be shifted due to  the presence of atoms. To
compensate for this shift,
we introduce a cavity detuning $\delta$ and numerically find $\delta=\delta_{\rm
opt}$ that  maximizes  either the efficiency $\eta$ or the product $\chi \eta$. 
The most dramatic improvements  
occur at lower $\Omega$,
as  shown by comparing Figs. \ref{fig:zero_delta} and \ref{fig:optim_eta}, and 
the dependence $\delta_{\rm opt}(\Delta,\Omega)$ is shown in Fig.
\ref{fig:optim_delta} for  $C\!=\!100$.
The optimum cavity detuning for large $\Omega$ is rather small, but  
it  further flattens the high efficiency ``plateau'' visible on Figs.
\ref{fig:zero_delta} and \ref{fig:optim_eta}.

Optimizing  $\chi \eta$ gives the upper homodyne detection limit, according to Eq. (\ref {chi-def}).
Ideally, the retrieval and detection efficiencies should be very close to each other
\begin{equation}\label{optim_func}
 {\rm max} [\chi(\delta,\omega_0)\eta(\delta)]\cong  {\rm max}[\eta(\delta)].
\end{equation}
Since the right hand-side of the last relation only depends on $\delta$, 
we can first find $\delta_{\rm opt}$ that optimizes
$\eta(\delta)$  and then subsequently optimize $\chi(\delta_{\rm
opt},\omega_0)\eta(\delta_{\rm opt})$ with respect to $\omega_0$. 
This optimization procedure can be  done very efficiently since  only one parameter
is varied at the time. This $\delta_{\rm opt}$ corresponds to the maximum
retrieval efficiency and thus it is an useful parameter in itself. 
The values  $\delta_{\rm opt}$ and $\omega_{\rm opt}$ obtained  in this optimization step 
can then be used as an initial
guess to find optimal $\delta$ and $\omega_0$ for $\chi \eta$. 
However this second optimization step  does not lead to any
practical increase in $\chi \eta$, as long as 
the linear phase approximation is valid (see Fig. \ref{fig:phase}). 
In that case, the optimization of $\chi \eta$ with respect to $\delta$ and
$\omega_0$ can be efficiently replaced by optimizing   $\eta(\delta)$ with
respect to $\delta$ and subsequently $\chi(\delta_{\rm opt},\omega_0)$  with
respect to $\omega_0$. 

For example, in the case presented in Fig.
(\ref{fig:phase})b (resp. (\ref{fig:phase})c), the $\chi \eta$-optimization  gives $\delta_{\rm opt}/2
\pi$ = 15.5 MHz (resp. -2.8  MHz) while the $\eta$-optimization gives $\delta_{\rm
opt}/2 \pi$ =15.7 MHz (resp. -5.2 MHz).  But even in the apparently  less favourable
case of Fig. (\ref{fig:phase})c,  the products  $\chi \eta$ are almost the same and nearly maximal, 
because the decrease in $\chi$ is compensated by an increase in $\eta$, so that 
$\chi \eta\!=\!0.930$ (resp. $\chi \eta\!=\!0.937$) as shown on Fig. \ref{fig:phase}.

\subsection{Calculation  of the efficiencies $\eta$ and $\chi \eta$}

Numerical results for $\eta$ and $\chi \eta$
are shown in Fig. \ref{fig:zero_delta} and   \ref{fig:optim_eta}  for
$\tau\!=\!150$ ns and $C\!=\!100$.  
On Fig. \ref{fig:zero_delta}  one has $\delta=0$ (i.e. no $\delta$-optimization
is performed), whereas Fig. \ref{fig:optim_eta} corresponds to  the optimized cavity
detuning $\delta_{\rm opt}(\Delta,\Omega)$, shown  on  Fig.
\ref{fig:optim_delta}. 
We observe that both 
quantities are
symmetric under the transformation $\Delta\rightarrow-\Delta$, consistently with
the  mappings (\ref{map1}) and (\ref{map2}).

Without cavity detuning optimization (see Fig. \ref{fig:zero_delta}), a plateau
of nearly maximum  retrieval efficiencies is reached for sufficiently large Rabi
frequencies $\Omega$ of the read field.  Increasing the pulse duration $\tau$
helps establishing the efficiency upper limit $C/(C\!+\!1)$ in a larger region
in the parameter space,  because the spectral distribution $|\eop(\omega)|$
becomes more narrow for longer pulses so that $\sim\omega^2/\kappa^2$ in Eq.
(\ref{P_int}) is negligible.
Increasing the cooperativity factor increases the efficiency limit $C/(C\!+\!1)$
but also widens the  plateau of nearly maximal efficiencies.
Since we keep $\kappa$ and $\gamma$ fixed, larger $C$ values imply larger
couplings $w$.
 However, for $\Delta\approx\delta\approx0$, the growth rate of $\eta$ as
$\Omega$ increases  is lower for larger $w$ because, according to the solution
(\ref{E-solut}), $\eop$ is  a function of the ratio $\Omega/\beta$ and
$\beta\sim w$.
Consequently, we need to operate with more intense read fields for larger $C$ in
order to retrieve single photons with near certainty. For example,
increasing $C$ from 100 to 200  for $\Delta\!=\!\delta=0$ requires $\sim40$ \%
higher Rabi frequencies to reach the plateau region.

\subsection{Optimized values of the efficiencies $\eta$ and $\chi \eta$}

In Fig. \ref{fig:optim_eta}, we show $\eta(\Delta,\Omega)$ assuming that
$\delta$ is optimized.
As expected, the cavity optimization yields to large improvements in efficiency
for smaller $\Omega$. As  Figs. \ref{fig:zero_delta}a and  \ref{fig:optim_eta}a
indicate, there is a region of relatively small  $\Delta$ and $\Omega$ in which
the $\delta$-optimization can significantly boost $\eta$ to reach almost 1, 
as shown for instance in Fig. \ref{fig:phase} a,b.

An overall effect of the $\delta$-optimization to $\chi \eta$ can be
seen by comparing  Figs. \ref{fig:zero_delta}b and  \ref{fig:optim_eta}b. 
As a general rule, the curve on  Figs. \ref{fig:optim_eta}b broadens and flattens, 
and high efficiency values become
available for lower readout Rabi frequencies  $\Omega$. 
 In general, increasing any of $C$ and $\tau$ expands the region of maximal $\chi \eta$. 
\begin{figure}[t]
\includegraphics[scale = 0.33]{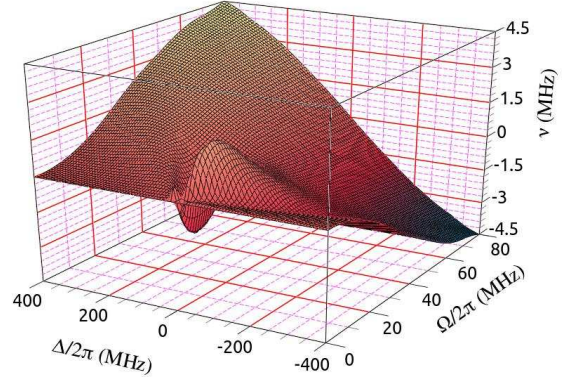}
\caption{(Color  online)
Frequency shifts $\nu\!=\!\omega_{\rm opt}/2\pi$
corresponding to optimized cavity detuning (the curve is antisymmetric with respect to the line $\Delta=0$, $\nu=0$).
These values of $\nu$
optimize the overall homodyne efficiency $\chi \eta$ in a linear phase
approximation (see Fig.~\ref{fig:optim_eta}). The parameters
are the same as in Fig.~\ref{fig:optim_delta}.
 \label{fig:optim_nu}
}
\end{figure}
Finally, Fig. \ref{fig:optim_nu} shows the  behaviour of $\nu_{\rm opt}=\omega_{\rm
opt}(\Delta,\Omega)/2\pi$.
The frequencies $\nu_{\rm opt}$ are rather small and of the order of a few MHz.
For $\Delta=0$ and real $\Omega$, $\eop(t)$ is real and negative
and thus no phase correction
is necessary. This could be a preferable
experimental choice, provided that a high enough $\Omega$ is available.   The
function
$\omega_{\rm opt}$ changes sign across the line $\Delta=0$ due to the symmetry
(\ref{map_omg}). As
previously pointed out, the  function $\nu_{\rm opt}$ 
exhibits distinguishable features for relatively small  $\Delta$ and $\Omega$
due to the cavity detuning optimization.

\section{Conclusion}

As a conclusion, we have calculated the amplitude and phase of a single-photon
wave-packet,  obtained by deterministically ``reading out'' a single polariton,
which  is assumed to be written beforehand onto a cloud of atoms within a
low-finesse optical cavity. This ``writing'' can be done either
non-deterministically by scattering a photon from the atoms  \cite{Vuletic}, or
deterministically by using Rydberg blockade techniques \cite{blockade}.  The
results indicate that both  the emission efficiency and the homodyne efficiency
towards an adequate local oscillator can reach values very close to one. This
opens the way towards the full quantum homodyne tomography of a single photon
deterministically generated within a single mode.

{\bf Acknowledgments} This work is supported by the ERC Advanced Grant ``DELPHI'' 
(Grant Agreement Number 246669) of the European Research Council. 

\section*{Appendix A}

As shown in ref. \cite{Gorshkov07},
the coupled propagation of  the field annihilation operator $\hat \eop(t)$ and
atomic operators $\hat P(t)$ and $\hat S(t)$ in the Heisenberg picture  writes~:
\begin{eqnarray}
\dot {\hat\eop} &=& -(\kappa + i \delta) \hat \eop + i w \hat P + \sqrt{2 \kappa} \;  \hat {\cal E}_{in}, \label{cav_eqs1-noise} \\
\dot {\hat P} &=& - (\gamma+i \Delta) \hat P + i w \hat \eop + i \Omega  \hat S + \sqrt{2 \gamma} \;  \hat {\cal F}_{P} , \label{cav_eqs2-noise}\\
\dot {\hat S} &=&  i \Omega^* \hat P , \label{cav_eqs3-noise}
\end{eqnarray}
where $\kappa$ and $\gamma$ are respectively the cavity and polarization decay
rates, while $\delta$ is the detuning between the  laser-driven Raman light
and the cavity, and $\Delta$  the detuning between the driving laser and the atomic line (see Fig.  \ref{fig:scheme}).   
Here we omit the spin-wave decay
rate, but it can be reintroduced
easily as shown in  \cite{Gorshkov07}.  The (bosonic vacuum) Langevin noise
terms are $\hat {\cal E}_{in}$ and
$\hat {\cal F}_{P}$.

In the article, we treated $\hat\eop$, $\hat P$ or $\hat S$ as complex numbers
and ignored all noise terms.
Here we want to explain what c-number quantity $\eop(t)$, $\ P(t)$ and $S(t)$ represent
and in which situations they can(not) be used. First, let us note that
$\eop(t)$, $\ P(t)$ and $S(t)$ are not expectation values of the corresponding
operators because
\begin{equation}\label{EPS_exp-vals}
\langle \hat\eop(t) \rangle= \langle \hat P(t) \rangle= \langle \hat S(t)  \rangle=0
\end{equation}
is fulfilled at all times. The last expression is true because
 there is only one excitation in the system, which may be in either of the
bosonic operators
$\hat\eop$, $\hat P$ or $\hat S $. Our main goal in this appendix is to show
that $\eop(t)$, $\ P(t)$ and $S(t)$ determine the expectation values of normally
ordered operators as follows
\begin{equation}\label{norm_exp-vals}
 \langle \hat A_i^{\dagger}(t) \hat A_j(t)\rangle=  A_i(t)^{*} A_j(t) ,
\end{equation}
where $\hat A_{i,j}$ can be any of the operators $\hat\eop$, $\hat P$ or $\hat S
$. We point out that the noise terms vanish in the propagation equations for
$\partial_t \langle \hat A_i^{\dagger}(t) \hat A_j(t)\rangle$ because we assume
that the input noise terms  $ \hat {\cal E}_{in}$ and $\hat {\cal F}_{P}$   are
in the vacuum state.
Then it is easy to check that all normally ordered noise correlations are zero
\cite{Gorshkov07}. This
means that, as far as the averages  $\langle \hat A_i^{\dagger}(t) \hat
A_j(t)\rangle$ are concerned, the propagation
equations for $\partial_t \langle \hat A_i^{\dagger}(t) \hat A_j(t)\rangle$ are
the same ones as for $\partial_t (A_i^*(t)  A_j(t))$, which can be obtained from
Eqs. (\ref{cav_eqs1-noise})-  (\ref{cav_eqs3-noise})  with all noise terms
set to zero.
Using this conclusion,  the relation (\ref{norm_exp-vals}) follows from the
following simple, but nonetheless important mathematical statement:
 If the evolution of operators $\hat A_i$  and  $\hat B_i$, $1\!\leq\! i
\!\leq\! n$ is linear
\begin{eqnarray}
\dot {\hat A}_i &=&\sum_s a_{is}(t)\hat A_s(t),  \label{Ai_DE},\\
\dot {\hat B}_i &=&\sum_s b_{is}(t)\hat B_s(t),  \label{Bi-DE}
\end{eqnarray}
and if the expectation values of all products $ \langle\hat B_i\hat A_j\rangle$
at some $t=t_0$ can be factorized as
\begin{equation}\label{AB_factor_init}
 \langle \hat B_i(t_0) \hat A_j(t_0)\rangle=B_{i0} A_{j0}
\end{equation}
then $ \langle\hat B_i(t)\hat A_j(t)\rangle$ are factorisable at all times
\begin{equation}\label{AB_factor}
 \langle \hat B_i(t) \hat A_j(t)\rangle=B_{i}(t) A_{j}(t) ,
\end{equation}
where $B_{i}(t)$ and $A_{j}(t)$  satisfy Eqs. (\ref{Ai_DE}) and (\ref{Bi-DE})
with the initial conditions $B_{i}(t_0)=B_{i0}$ and $A_{i}(t_0)=A_{i0}$.

We can directly check that the products $B_{i}(t) A_{j}(t)$  always satisfy the
equation of motion for $\langle \hat B_i(t) \hat A_j(t)\rangle$. However, to  be
solutions for $\langle \hat B_i(t) \hat A_j(t)\rangle$, the   products $B_{i}(t)
A_{j}(t)$ have to also satisfy the initial conditions. This is only possible if
all $\langle \hat B_i(t_0) \hat A_j(t_0)\rangle$ are factorisable as required by
 Eq. (\ref{AB_factor_init}). We emphasize that this factorization of $\langle \hat
B_i(t_0) \hat A_j(t_0)\rangle$ is not a trivial condition for operators and it
is not generally fulfilled.

In our case, the only initial nonzero expectation value is
 $\langle\hat S^{\dagger}(t_0)\hat S(t_0)\rangle$ so the initial averages are clearly
factorisable. This means that the condition (\ref{AB_factor_init}) is  fulfilled and so
is the relation  (\ref{norm_exp-vals}).  Therefore, we have shown that
the expectation values of the normally ordered operators are exactly the same as
for the products of the c-number quantities $\eop (t)$,  $P(t) $ and $S(t) $.
For simplicity, $S(t_0)=1$ is used  because the only effect  of  another phase
convention  $S(t_0)\rightarrow S(t_0) e^{i\phi_0}$ is that all $\eop (t)$, $P(t) $,
and  $S(t) $ acquire the same additional  phase factor $ e^{i\phi_0}$ which leaves
the averages (\ref{norm_exp-vals}) unchanged.

In case where there is only one excitation in the system, the condition (\ref{AB_factor_init}) is 
equivalent to the requirement that  the initial state is a pure state.   To show this, we consider here the evolution of the density matrix
of the system (in the Schr\"odinger picture), rather than the evolution of the operators (in the Heisenberg picture) used up to now.
We take as a basis the Fock states  $\{ \left|000\right>, \, \left|100\right>, \, \left|010\right>,  \, \left|001\right> \}$ for
the bosonic operators $\hat\eop$, $\hat P$ or $\hat S $, where the states are labeled as $\left|n_{\eop} \, n_P \, n_S\right>$.
The state $\left|000\right>$ does not evolve under the Hamiltonian  $H$ of the considered three-level atoms ($H\left|000\right>=0$),
but only due to damping, and
the Hamiltonian part of the evolution takes place within the subspace spanned by $\left|100\right>$, $\left|010\right>$, and $\left|001\right>$.
Realistic atoms may require considering more states, but in our problem the calculation of this $3\times3$ 
sub-matrix is equivalent to the calculation of  $\eop (t)$,  $P(t) $ and $S(t) $ as carried out  in  this paper.  Denoting $\{\hat A_i\}=\{\hat\eop,\hat P,\hat S\}$,
the density matrix elements are simply
\begin{equation}\label{r_ij}
 r_{ij}=\langle \hat A_i^{\dagger} \hat A_j\rangle(t).
\end{equation}
If the factorization condition (\ref{AB_factor_init}) is satisfied, then
$r_{ij}(t_0)=A_i^{*}(t_0)A_j(t_0)$, which implies  $\hat \rho(t_0)$ is a projector, and
therefore that  the initial density matrix describes a pure state.
The other implication stating that if the initial state is pure then Eq. (\ref{AB_factor_init}) is satisfied follows directly from the  definition  (\ref{r_ij}).
Even more importantly, due to Eq. (\ref{AB_factor}), the factorization  is preserved at all times which implies that the density matrix
 takes the form
\begin{equation}\label{r_ij-time}
\hat\rho(t)= r(t)  |000\rangle \langle 000 | +   \left|\psi(t)\right>\left<\psi(t)\right|,
\end{equation}
where $\left|\psi(t)\right>=(\eop(t),P(t),S(t))^T$    is a (non-normalized) pure state.
Therefore (up to a normalizing factor)  the initial state  $\left|001\right>$ evolves into the superposition  
$\eop(t)\left|100\right> + P(t)\left|010\right>+S(t)\left|001\right>$. We see that the component of the single-photon state is exactly $\eop(t)$.

\section*{Appendix B}

We want to confirm Eq. (\ref{bad_cav-corr}) which asserts that, for sufficiently
large pulsewidths $\tau$, the quantum field $\eop(t)$ simply follows $P(t)$
\begin{equation}\label{retard_e}
 \eop(t)= i w P(t-\delta t) /\kappa
\end{equation}
with a constant time delay $\delta t=1/\kappa$. This relation is very similar to
the relation between $\eop(t)$ and $P(t)$ in the bad cavity limit $\kappa\gg w$
(which is not fulfilled in our case) except for the delay $\delta t$. This
similarity is related to the relation (\ref{bad_cav_abs-val}) in the frequency
domain. In the time domain, we will show that it is closely related to the fact
that the single photon escapes the cavity in a very early stage of the retrieval
sequence.

Let us assume that the read pulses are parameterized  as $\Omega(t)=\Omega_0
f(t/\tau)$. By using the dimensionless quantities $t/\tau$,  $k\tau$, $\ldots$,
the number of parameters in Eqs. (\ref{cav_eqs1})-(\ref{cav_eqs3}) is
effectively reduced by one, which makes the scaled quantities  more natural.
We can interpret Eq. (\ref{bad_cav-corr}) as the definition of
a new function $\delta t(t)$. If  $\delta t(t)/\tau$ is small  at all times,
then the conclusion  $\delta t=1/\kappa$ directly  follows that from Eqs.
(\ref{cav_eqs1}) and (\ref{bad_cav-corr}). Therefore, we only need to show that
$\delta t(t)/\tau$ is indeed small at all times for sufficiently long read
pulses. This can be explicitly shown using the solution (\ref{E-solut}).  The
implication that $1/\kappa$ becomes a short time scale in this limit also
indicates that the single photon has been emitted very early in terms of
relative times $u=t/\tau$.

For long pulses, the real part of the argument of the exponential function in
Eq. (\ref{E-solut}) acquires very large negative values which indicates that the
photon emission has been terminated.
The time interval during which the photon has been emitted can be estimated from
the condition
\begin{equation} \label{emission-interv}
{\rm Re}\left[\kappa \tau \int_{u_0}^u
\frac{\Omega_0^2f^2(u')}{\beta^2+\Omega_0^2f^2(u')} du' \right] \approx 1.
\end{equation}
Therefore, increasing $\tau$ at fixed $\Omega_0$ results in a photon emission at
shorter relative times $u={t/\tau}$.  In the limit of very long pulses, all
terms $\sim{\dot \Omega}/\Omega$ are negligible.  Consequently, according to Eq.
(\ref{E-DE}),
\begin{equation} \label{E-ratio}
\frac{1}{\eop}  \frac{d{\eop} }{dt}  \approx
\frac{|\Omega|^2}{\beta^2+|\Omega|^2} \rightarrow 0
\end{equation}
holds at all times during the photon emission.
 Combining Eqs. (\ref{bad_cav-corr}), (\ref{cav_eqs1}), and (\ref{E-ratio}), we
get that
\begin{equation} \label{E-ratio2}
\left( \eop(u+\delta t(t)/\tau) -\eop(u)\right) / \eop(u) \rightarrow 0
\end{equation}
is satisfied for all $t$  during the photon emission. There are two logical
possibilities consistent with the last equation: either $\delta t(t)/\tau$ is
small at all times or $\delta t(t)$ relates distant instants  with the same
$|\eop|$ and phase $\theta_{\eop}$ of the quantum field.  The latter possibility
is very restrictive and would be essentially an additional  condition to the
equations  of motion (\ref{cav_eqs1})-(\ref{cav_eqs3}) since such distant
correlations cannot naturally come from strictly local propagation equations. In
fact, all panels in Fig. \ref{fig:phase} clearly show that the points with the
same
$|\eop|$ have different phases so that the only consistent possibility derived
from  Eq. (\ref{E-ratio2}) is that  $\delta t(t)/\tau$ is small for all $t$ and
consequently  $\delta t=1/\kappa$.



\end{document}